\documentclass{latex/webofc}
\input{latex/packages.tex}

\newcommand*{\TeV}{\ensuremath{\text{Te\kern -0.1em V}}}
\newcommand*{\GeV}{\ensuremath{\text{Ge\kern -0.1em V}}}
\newcommand*{\MeV}{\ensuremath{\text{Me\kern -0.1em V}}}
\newcommand*{\keV}{\ensuremath{\text{ke\kern -0.1em V}}}
\newcommand*{\eV}{\ensuremath{\text{e\kern -0.1em V}}}

\begin{document}

\title{Software Citation in HEP: Current State and Recommendations for the Future}

\author{\firstname{Matthew} \lastname{Feickert}\,\orcidlink{0000-0003-4124-7862}\inst{1}\fnsep\thanks{Corresponding author \email{matthew.feickert@cern.ch}} \and
 \firstname{Daniel} \lastname{S. Katz}\,\orcidlink{0000-0001-5934-7525}\inst{2} \and
 \firstname{Mark} \lastname{S. Neubauer}\,\orcidlink{0000-0001-8434-9274}\inst{2} \and
 \\
 \firstname{Elizabeth} \lastname{Sexton-Kennedy}\,\orcidlink{0000-0001-9171-1980}\inst{3} \and
 \firstname{Graeme} \lastname{A. Stewart}\,\orcidlink{0000-0003-0182-7088}\inst{4}
}

\institute{University of Wisconsin-Madison, Madison, Wisconsin, USA
 \and
 University of Illinois Urbana-Champaign, Illinois, USA
 \and
 Fermi National Accelerator Laboratory, Illinois, USA
 \and
 CERN, Switzerland
}

\abstract{%
 In November 2022, the HEP Software Foundation and the Institute for Research and Innovation for Software in High-Energy Physics organized a workshop on the topic of Software Citation and Recognition in HEP.
The goal of the workshop was to bring together different types of stakeholders whose roles relate to software citation, and the associated credit it provides, in order to engage the community in a discussion on: the ways HEP experiments handle citation of software, recognition for software efforts that enable physics results disseminated to the public, and how the scholarly publishing ecosystem supports these activities.
Reports were given from the publication board leadership of the ATLAS, CMS, and LHCb experiments and HEP open source software community organizations (ROOT, Scikit-HEP, MCnet), and perspectives were given from publishers (Elsevier, JOSS) and related tool providers (INSPIRE, Zenodo).
This paper summarizes key findings and recommendations from the workshop as presented at the 26th International Conference on Computing in High Energy and Nuclear Physics (CHEP 2023).

}
\maketitle

\section{Introduction}\label{sec:introduction}
Software is a research product --- an asset created as a byproduct of scientific research --- that is ubiquitously used in and necessary to physics research, though it is not always given the same levels of importance and scholarly weight as other research products like publications and data products~\cite{Cranmer:2021urp}.
In November 2022, the HEP Software Foundation (HSF) and the Institute for Research and Innovation for Software in High-Energy Physics (IRIS-HEP)~\cite{S2I2HEPSP,IRISHEPWEB} organized a topical workshop on software citation and recognition in the field of high energy physics (HEP)~\cite{software_citation_workshop_report,software_citation_indico}.
The goal of the workshop was to provide a community discussion around ways in which HEP experiments handle citation of software and recognition for software efforts that enable physics results disseminated to the public.
The workshop participants and primary presentations were from the LHC experiments that are primary stakeholders in IRIS-HEP operations: ATLAS, CMS, and LHCb; the particle physics open source software development communities: ROOT Team, Scikit-HEP~\cite{Rodrigues:2020syo}, MCnet, and IRIS-HEP; as well as the scientific publishing community and ecosystem most involved with HEP: Elsevier, the Journal of Open Source Software (JOSS)~\cite{smith_journal_2018}, and INSPIRE~\cite{INSPIRE}.

The principles of software citation that the HEP community is interested in engaging with are those established by the FORCE11 Software Citation working group~\cite{smith_software_2016}.
These principles are defined as:

\begin{enumerate}
    \item \textbf{Importance}: Software should be considered a legitimate and citable product of research.
Software citations should be accorded the same importance in the scholarly record as citations of other research products, such as publications and data; they should be included in the metadata of the citing work, for example in the reference list of a journal article, and should not be omitted or separated.
Software should be cited on the same basis as any other research product such as a paper or a book, that is, authors should cite the appropriate set of software products just as they cite the appropriate set of papers.
    \item \textbf{Credit and Attribution}: Software citations should facilitate giving scholarly credit and normative, legal attribution to all contributors to the software, recognizing that a single style or mechanism of attribution may not be applicable to all software.
    \item \textbf{Unique Identification}: A software citation should include a method for identification that is machine actionable, globally unique, interoperable, and recognized by at least a community of the corresponding domain experts, and preferably by general public researchers.
    \item \textbf{Persistence}: Unique identifiers and metadata describing the software and its disposition should persist --- even beyond the lifespan of the software they describe.
    \item \textbf{Accessibility}: Software citations should facilitate access to the software itself and to its associated metadata, documentation, data, and other materials necessary for both humans and machines to make informed use of the referenced software.
    \item \textbf{Specificity}: Software citations should facilitate identification of, and access to, the specific version of software that was used.
Software identification should be as specific as necessary, such as using version numbers, revision numbers, or variants such as platforms.
\end{enumerate}

Today the global research community now has these principles, citation policies from journal publishers, and modern open source tooling to facilitate the generation of software citations.
There has also been growing movement among research software developers, research paper authors, and journal reviewers and editors~\cite{smith_journal_2018} towards an increase in software citation.
For the HEP community it is important to understand the current state (as of 2023) of software citation norms and culture in the field and how its importance can be conveyed and supported through community tooling, standards, and practices.

\section{Current State of Software Citation in HEP}\label{sec:current_state}

\subsection{LHC Experiments}\label{sec:lhc_experiments}

To understand the current state of software citation in the field reports from the ATLAS, CMS, and LHCb experiments were given that summarized the experiments' current standards and practices and future plans.
ATLAS takes the approach of using a ``catch-all'' citation of all ATLAS software and firmware through the citation of an ATLAS public note that ``briefly describes the software and provides links to dynamic and persistent repositories wherein the code resides''~\cite{ATL-SOFT-PUB-2021-001}.
This public note is then cited in many ATLAS papers.
ATLAS additionally cites the paper for the ATLAS detector simulation software~\cite{SOFT-2010-01} as well as GEANT4~\cite{GEANT4:2002zbu}, and the Monte Carlo simulation generators used~\cite{Sjostrand:2007gs,Sjostrand:2014zea,Alwall:2014hca,Sherpa:2019gpd}.
In terms of statistical analysis ATLAS cites the methodology papers that describe the techniques used in analyses, but in general does not cite the actual software that implements the techniques, with the notable exception of machine learning libraries~\cite{chollet2015keras,tensorflow2015-whitepaper}.
Citation practices are not uniformly consistent in the experiment though, with some physics groups beginning to regularly cite statistical libraries that provide clear citation guidelines~\cite{pyhf,pyhf_joss} (Principles 1 and 2).

CMS similarly has an established culture of regularly and consistently citing Monte Carlo generators, GEANT4, and machine learning tools.
However, they note there could be improvement in the citation of the software that CMS itself produces, both in experimental internal notes and documentation as well as scientific publications.
CMS also expressed positive views towards starting practices of publishing papers --- either as CMS Collaboration publications or as limited authorship papers from the CMS Software and Computing Group --- on CMS software, bringing with it increased visibility of scientific software development, documentation standards, and references of software version information (Principles 1 and 2).

LHCb has taken a more proactive stance on software citation following recommendations presented at the CHEP 2018 Conference~\cite{CHEP-2018-recommendations} by providing an internal LHCb software citation starting template for software commonly used in analysis.
Analysis teams are then encouraged to revise the template with the citations of the software used in their analysis with the goal that all high-level software used is properly cited (Principles 1, 2, and 6).
These practices are encouraged in the collaboration, but not explicitly required, and so analysis teams may require citation guidelines to be provided.
LHCb also noted that the citation practices of the HEP community are largely due to cultural norms rather than technical challenges, and that while LHCb strives to be citing more software in the future having LHC community recommendations on software citation would be useful for motivating better practices.

\subsection{Software Projects}\label{sec:software_projects}

Views from prominent open source software projects and software communities inside of HEP were also discussed, with a broad range of community cultural views and practices.
The ROOT team noted they explicitly are not interested in ROOT's software citation, as the ROOT team does not view it as adding value to their work, that updating citation information would require additional effort, and in the team's view the current HEP culture of citation with journal publications for larger software projects is working well.
The ROOT team was careful to note though that these views are specifically limited to software citation for ROOT~\cite{Brun:1997pa} and should not be viewed as being universal.
In contrast, the Scikit-HEP community project has prioritized adopting software citation recommendations and tooling from the broader scientific open source community (e.g. Zenodo~\cite{zenodo}, \texttt{CITATION.cff} files~\cite{Druskat_Citation_File_Format_2021}) to provide credit to the developers producing community tools (Principle 2) as well as recognize project contributions of multiple types~\cite{all-contributors}.
Scikit-HEP views software citation as important to their community and would welcome HEP community guidelines to guide users of the community tools to easily and correctly cite the software.
The MCnet community noted that as a community of Monte Carlo generator software projects they have benefited from consistent citation by the LHC experiments.
Several community factors lead to this culture, including the MCnet community becoming organized in the leadup to the start of the LHC and providing clear citation guidelines and often making programmatic citation information available from the software itself.
MCnet raised the potential problems with the current citation model of citing papers for large releases of the software as this does not equally value or reward the development and maintenance labor that occurs between the long intervals between publications.
As a result, MCnet is interested in both technical solutions as well as community guidelines and policy regarding software citation.

\subsection{Publishing Community}\label{sec:publishers}

Following the state of software citation in the HEP community, views and recommendations from INSPIRE, Elsevier, and JOSS were shared given their different roles related to scientific publishing and citation.
INSPIRE is an integral part of how HEP interacts with publications, related metadata, and acquires updated citation information as tracked submissions move from preprints to publication.
Having these capabilities for the citation information for software in HEP would be a technical boon.
While INSPIRE currently only handles software papers, there are plans to add support for data products and software in the future, initially by harvesting metadata from relevant trusted repositories (e.g. INSPIRE HEP Zenodo community, HEPData, CERN Open Data portal).
This information would be gathered by software digital object identifier (DOI), and could be aggregated across multiple releases of the same software.
It is therefore important that software projects that seek citations in the future provide DOIs now (Principles 3, 5, and 6).
Elsevier noted that it is the responsibility of the scientific community to reach a consensus on how to cite software and to share these guidelines with publishers, which can then better instruct journal editors and referees what the expectations for citation are and how to support them.
JOSS noted that in addition to incentivizing high quality research software with the journal guidelines and review standards, JOSS can also help bridge the cultural and technical gaps between traditional publication citation and the citation of software directly.

\section{Recommendations}\label{sec:recommendations}

In addition to establishing guidelines for the HEP community, providing recommendations of software citation best practices and supported tooling aids in community adoption of new guidelines.
A behavior step that can be implemented is for software projects to clearly document a recommended citation and have this information be easily findable anywhere the software source code or distributions are hosted or documented (e.g. version control repositories, public documentation websites, package indexes, archives).
There has been historical precedent in HEP for tools to provide recommendations for how to cite the software being used by printing it as a runtime banner to standard output, as seen in~\Cref{lst:pythia_banner}.
This method was developed before citation conventions were established more broadly in the scientific computing community, and modern practices would generally avoid interrupting user logs with this information.
It is instead preferable, in addition to having a clearly documented and advertised recommended citation, to provide citation APIs in the software --- both at the language level and at the command line interface if the software supports one.

In addition to having clear citation recommendations, it is beneficial to adopt a standardized citation file format.
A strong choice is the recent Citation File Format~\cite{Druskat_Citation_File_Format_2021} which is serialized as YAML as a \texttt{CITATION.cff}, as seen in~\Cref{lst:cation.cff}.
\texttt{CITATION.cff} files have the benefit of being both human- and machine-readable with a well defined, versioned schema.
Through related tooling \texttt{CITATION.cff} can also be programmatically validated against schemas and converted to other citation formats (e.g., BibTeX, CodeMeta, EndNote, RIS, schema.org, Zenodo, APA).
\texttt{CITATION.cff} also benefits through supported integration with GitHub\footnote{Providing a ``Cite this repository'' button on a repository with a \texttt{CITATION.cff} file.}, Zenodo, and Zotero, allowing for the citation information to be reliably exported to multiple services from a single file (Principle 2).
The integration with Zenodo is significant, as the HEP community is already frequent users of Zenodo for long term archival of source code (Principle 4) and DOI generation for the source code of software releases (Principle 3).
Software projects that adopt the use of \texttt{CITATION.cff} and archive the source code with Zenodo have a clearly defined toolchain provenance for citation information dissemination (Principle 5).
Given this, it is recommended that there is a single source of truth for citation information, such as a \texttt{CITATION.cff} file, that is under version control with the software source code and is used to generate all other metadata or forms of citation information by other services.

\begin{lstlisting}[
    basicstyle=\ttfamily\tiny,
    float=tp,
    floatplacement=tbp,
    caption={Sections of an example runtime banner printed to standard output from \texttt{PYTHIA 8.2}~\cite{Sjostrand:2014zea} with citation guidelines --- a historical community norm.},
    captionpos=b,
    label={lst:pythia_banner}
]
                  *------------------------------------------------------------------------------------*
                  |                                                                                    |
                  |  *------------------------------------------------------------------------------*  |
                  |  |                                                                              |  |
                  |  |                                                                              |  |
                  |  |   PPP   Y   Y  TTTTT  H   H  III    A      Welcome to the Lund Monte Carlo!  |  |
                  |  |   P  P   Y Y     T    H   H   I    A A     This is PYTHIA version 8.230      |  |
                  |  |   PPP     Y      T    HHHHH   I   AAAAA    Last date of change:  6 Oct 2017  |  |
                  |  |   P       Y      T    H   H   I   A   A                                      |  |
                  |  |   P       Y      T    H   H  III  A   A    Now is 06 May 2023 at 01:12:28    |  |
                  |  |                                                                              |  |
                  ...
                  |  |   The main program reference is 'An Introduction to PYTHIA 8.2',             |  |
                  |  |   T. Sjostrand et al, Comput. Phys. Commun. 191 (2015) 159                   |  |
                  |  |   [arXiv:1410.3012 [hep-ph]]                                                 |  |
                  |  |                                                                              |  |
                  |  |   The main physics reference is the 'PYTHIA 6.4 Physics and Manual',         |  |
                  |  |   T. Sjostrand, S. Mrenna and P. Skands, JHEP05 (2006) 026 [hep-ph/0603175]  |  |
                  |  |                                                                              |  |
                  |  |   An archive of program versions and documentation is found on the web:      |  |
                  |  |   http://www.thep.lu.se/Pythia                                               |  |
                  |  |                                                                              |  |
                  |  |   This program is released under the GNU General Public Licence version 2.   |  |
                  |  |   Please respect the MCnet Guidelines for Event Generator Authors and Users. |  |
                  |  |                                                                              |  |
                  ...
\end{lstlisting}

\setcounter{listing}{1}
\begin{listing}
\begin{minted}{yaml}
cff-version: 1.2.0
message: "If you use this software, please cite it as below."
authors:
  - family-names: Druskat
    given-names: Stephan
    orcid: https://orcid.org/0000-0003-4925-7248
title: "My Research Software"
version: 2.0.4
doi: 10.5281/zenodo.1234
date-released: 2021-08-11
\end{minted}
 \caption{Example of a minimal \texttt{CITATON.cff} file using v1.2.0 of the CFF schema~\cite{Druskat_Citation_File_Format_2021}.}
 \label{lst:cation.cff}
\end{listing}

\section{Conclusions}\label{sec:conclusions}

Revisiting the software citation principles in the view of current approaches and technologies in HEP provides a structure for starting community guidelines:

\begin{enumerate}
    \item \textbf{Importance}: As a field HEP understands software is important, but improvements could be made on views towards software as a research product.
    \item \textbf{Credit and Attribution}: The giving of credit is improving in HEP, but the community can leverage software friendly journals (e.g., JOSS) to help accelerate this.
    \item \textbf{Unique Identification}: Use of Zenodo archives already exists in HEP, which provides well integrated tooling for DOI generation.
The use of \texttt{CITATION.cff} files in software repositories can help as well.
    \item \textbf{Persistence}: Zenodo provides long term archival of source code and project metadata.
    \item \textbf{Accessibility}: HEP is becoming more FAIR~\cite{wilkinson_fair_2016,chue_hong_neil_p_2022_6623556} focused, bringing with it an increased focus on accessibility.
As \texttt{CITATION.cff} provides a common framework for metadata, adopting it as a community standard for software citation information allows for greater accommodation and discovery by citation discovery tools.
    \item \textbf{Specificity}: Version numbers of software should be included in \texttt{CITATION.cff} files and the version used for analysis should be reported in publications.
\end{enumerate}

It is seen there are both social and technical tooling challenges to be addressed to reach HEP community guidelines and recommendations for software citation.
While there exist multiple practices towards software citation in the HEP community today, this should not be viewed as a large challenge towards global community standards adoption as variations in homogeneity of practice are common even in journal publication.
The community wide agreement that software citation is important, should be practiced more often, and provides both social and technical benefits gives sufficient motivation to develop HEP community wide recommendations in the near future.

\section*{Acknowledgments}\label{sec:acknowledgments}
Matthew Feickert and Daniel S. Katz are supported by the U.S. National Science Foundation (NSF) under Cooperative Agreement OAC-1836650 (IRIS-HEP).
Mark S. Neubauer is supported by the U.S. Department of Energy, Office of Science, High Energy Physics, under contract number DE-SC0023365, and by the National Science Foundation under Cooperative Agreement OAC-1836650 (IRIS-HEP).

\bibliography{bib/ref}

\end{document}